# The Emotional Impact of Game Duration: A Framework for Understanding Player Emotions in Extended Gameplay Sessions

Anoop Kumar[1], Suresh Dodda[2], Navin Kamuni[3], Venkata Sai Mahesh Vuppalapati[4]

[1]Anoop.kumar.2612@gmail.com, [2]sureshr.dodda@gmail.com, [3]navin.kamuni@gmail.com,
[4]mahesh26sai@gmail.com

[1]IIT Roorkee, India
[2]IT Department, Eudoxia Research center, USA
[3]AI-ML, BITS Pilani WILP, USA,
[4]Tubi TV, USA

*Abstract*—Video games have played a crucial role in entertainment since their development in the 1970s, becoming even more prominent during the lockdown period when people were looking for ways to entertain them. However, at that time, players were unaware of the significant impact that playtime could have on their feelings. This has made it challenging for designers and developers to create new games since they have to control the emotional impact that these games will take on players. Thus, the purpose of this study is to look at how a player's emotions are affected by the duration of the game. In order to achieve this goal, a framework for emotion detection is created. According to the experiment's results, the volunteers' general ability to express emotions increased from 20 to 60 minutes. In comparison to shorter gameplay sessions, the experiment found that extended gameplay sessions did significantly affect the player's emotions. According to the results, it was recommended that in order to lessen the potential emotional impact that playing computer and video games may have in the future, game producers should think about creating shorter, entertaining games.

## I. INTRODUCTION

Video games have had a tremendous impact on our lives and the entertainment industry over the past 50 years. Games and gaming have become much more popular because of the invention and development of video games since the 1970s. This has been particularly common during the lockdown, when many people looked for comfort in entertainment. A Statista report [1]–[6] claims that during the pandemic from March 16, 2020 to March 22, 2021—the overall number of games sold increased by 63% globally, reaching 4.3 million. The fact that like-for-like sales increased overall and exceeded 44% serves as additional support for this. Video games are primarily computer games made with the intention of entertaining players. They are interactive and provide access to 2D and 3D virtual worlds that are not possible in the actual world. Based on the game's genre and media type, these worlds—also referred to as virtual environments—will follow specific laws and regulations [7]–[13]. Video games are available on a variety of platforms, including consoles, PCs, and smartphones. A wide range of game genres are available, including party games, trivia, fighting games, and survival games.

The purpose of this study is to provide light on how gameplay influences players' emotions and engagement by examining the effect of game duration on players' emotional states. It looks at how players' emotions change while they play games in an effort to give creators insights into how to make more engaging experiences that suit player preferences and feelings. This research will serve as a foundation for future game design studies.

## II. STATE OF THE ART

This study explores different approaches to quantify players' emotional responses in real time while examining the emotional impact of video games. Prior studies have focused on the importance of emotional responses, particularly while playing violent video games. Using the chilli sauce paradigm, participants' propensity to distribute different amounts of chilli sauce after playing violent versus non-violent games was used as an indirect indicator of their levels of aggression. This is only one creative technique that was emphasized. This technique provided a novel tool to measure emotional states, demonstrating increased levels of rage connected to violent video games. Though participants may have played games like "Call of Duty" before, the choice of these titles for the trials could introduce bias and magnify emotional responses in ways that might not be fully representative of the game's influence on a novice player. This component highlights how difficult it can be to precisely measure emotional responses in gaming environments and raises the possibility that carefully chosen games should be used in these kinds of investigations. Other study techniques have involved observing participants' eye movements to deduce their emotional states. Results have shown that playing violent video games is associated with higher levels of annoyance and fury as well as less frequent blinking. While it also emphasizes the drawbacks of depending just on physiological responses to infer emotional states, this association between eye movements and emotional reactions broadens our understanding of how players respond to game content. Using a different methodology, research varied the degree of control players had over a custom-developed tower defense game to investigate players' perceptions of accomplishment and control. The results indicated the importance of control as a component determining gaming experience, with higher levels of perceived control appearing to increase players' emotional well-being. In order to gather information on the emotional states of the players, this study used questionnaires, which provided direct participant insights but also limited the range of emotion detection to self-reported measurements. Some researchers have shown that using questionnaires as the only means of obtaining emotional data suggests that there may be a gap in the collection of all the emotions that players can feel. Although they offer insightful subjective information, the lack of further, objective emotion detection technologies may restrict the extent of analysis that can be conducted. One tool that could provide a more detailed understanding of players' emotional reactions is facial expression detection, which is conspicuously lacking in the evaluated studies. In addition, the duration of games appears to be an important but little-studied element influencing emotional reactions. This flaw points to a potential avenue for future research that examines the ways in which varying game exposure durations affect players' emotional states. The current work intends to create a more comprehensive method of emotion recognition in gaming by utilizing these insights. Through the integration of cutting-edge techniques such as the chilli sauce paradigm with sophisticated emotion detection technology, the study aims to provide a thorough understanding of how emotions change throughout different game durations. This multidisciplinary approach offers new opportunities for research into the intricate interactions between technology, human behavior, and emotion, and it promises to improve our understanding of the emotional environment in digital gaming.

### III. MATERIALS AND METHODS

The study aims to combine technology innovation with psychological knowledge in order to develop an emotion recognition program. The study was primarily motivated by the goal of using the flexible and commonly available Python programming language to detect and measure emotional reactions in real-time. Given its ease of use, adaptability, and remarkable statistical data processing capabilities—a crucial aspect for the study's testing phase—Python was chosen strategically. The first step in the process was creating and coding a rudimentary (refer to Fig. 1) to access and communicate with video recording devices, including internal or external cameras. Establishing a direct channel of communication with the participants required taking this step. The application was able to automatically establish connections with cameras by using the OpenCV[1] library. The program could also adjust to different hardware configurations by instructing the user and providing an error message. After the camera was successfully integrated, the application proceeded to record and manipulate video streams, presenting them to the user while continuously keeping an eye out for a certain command to "exit". Face detection is a more sophisticated process, and this degree of interaction laid the groundwork for it. By using a Haar Cascade classifier[2], the software was able to recognize faces in the video stream. This machine learning-based technique analyzed pre-trained models on both positive (faces present) and negative (faces absent) images to enable effective object detection. Accurate face detection was made easier by this technique, which also prepared the way for the later, more complex stage of emotion recognition.

The study's primary focus was on using the DeepFace[3] framework to identify and analyze emotions. This approach was chosen because of the framework's track record of accuracy and efficiency in facial analysis in a variety of circumstances, including places with poor lighting. With the addition of this framework, the program made a major advancement in its ability to accurately identify a range of emotions. The SelfiSegmentation[4] class from the cvzone[5] module was used to improve concentration and reduce background noise, which improved the program's analytical powers even more. The study included more features for user interaction and data management in order to get ready for the trial phase. Carefully considering all the factors that could affect the results, the experimental design was created to explore the connection between playtime length and emotional response. This included the game selection, the experiment location, and the lighting setup, all of which were taken into account and managed to guarantee the accuracy of the data gathered. A group of volunteers was chosen from a determined group of players, and they played the recently published game SplitGate[6] in sessions that lasted different lengths. They were chosen from an eager group of gamers (i.e. authors of this paper). This configuration offered a singular chance to track the evolution of emotional dynamics, revealing insights into the potential effects of prolonged exposure to digital entertainment on emotional states. Strict data gathering procedures, anonymization, and encryption were used to guarantee that moral principles were respected. By utilizing advanced programming methods, machine learning algorithms, and careful experiment planning, it shed light on the intricate relationship between technology, human behavior, and emotion. The knowledge gained from this study not only advances the subject of human-computer interaction but also creates new opportunities for research into emotional intelligence and its applications in virtual settings.

---

[1] https://opencv.org/
[2] https://docs.opencv.org/3.4/db/d28/tutorial_cascade_classifier.html
[3] https://github.com/serengil/deepface
[4] https://github.com/yasirrustam06/SelfiSegmentation
[5] https://github.com/cvzone/cvzone
[6] https://www.splitgate.com/

The flowchart of the pre-methodology is presented in Fig. 1.

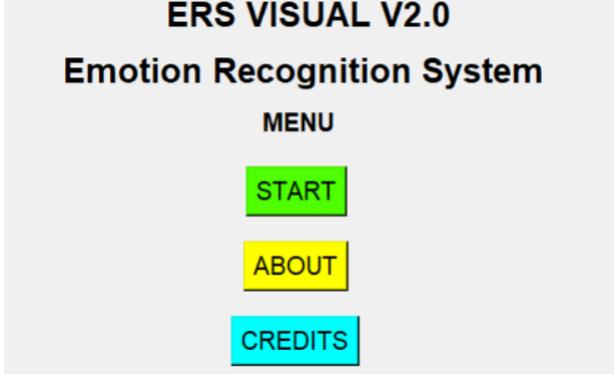

Fig. 1. The outcome of creating a generic menu for the emotion detecting application with Tkinter

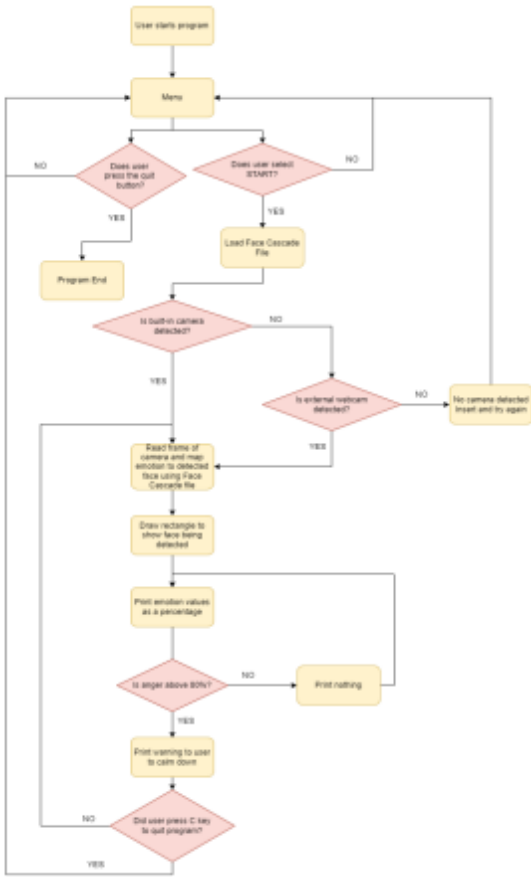

Fig. 2. This flowchart illustrates the program's flow as well as the design process for each of its component parts

## IV. EXPERIMENTAL ANALYSIS

Using an emotion detection program to track seven different emotions, the study examined how players' moods changed over the course of 20, 40, and 60-minute gaming sessions. Early sessions revealed a high level of grief and dread mixed with some happiness. Extended sessions led to a decrease in happiness and an increase in fear, hostility, and disgust. Extended gameplay resulted in increased negative affect and reduced enjoyment, underscoring the intricate emotional dynamics present in video games.

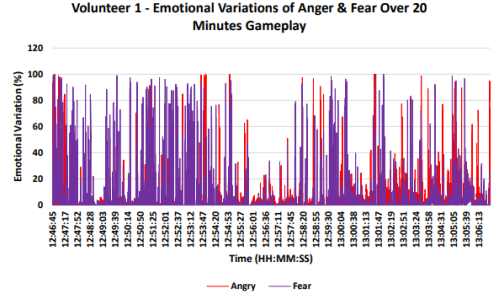

Fig. 3. Volunteer 1: Anger and Fear's Emotional Variations During 20 Minutes of Play

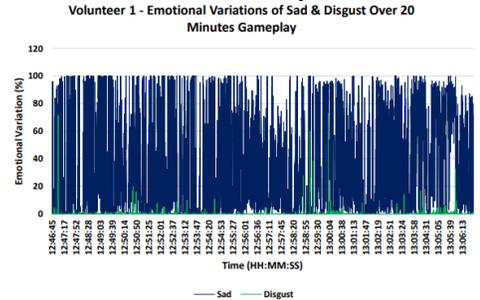

Fig. 4. Volunteer 1: Sad and Disgust Emotional Variations During 20 Minutes of Play

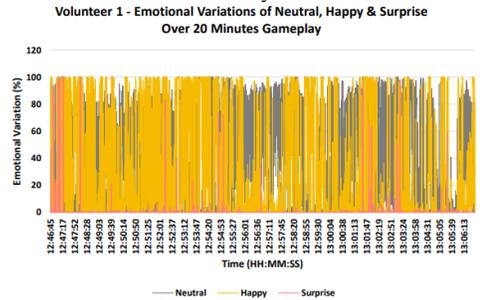

Fig. 5. Volunteer 1: Neutral, Happy, and Surprise Emotional Variations During 20 Minutes of Play

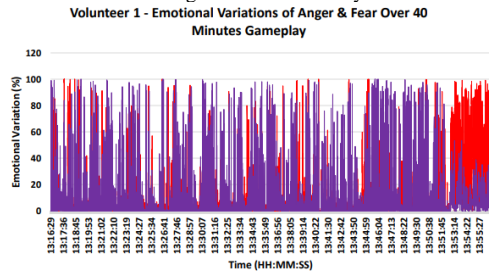

Fig. 6. Volunteer 1: Anger and Fear's Emotional Variations During 40 Minutes of Play

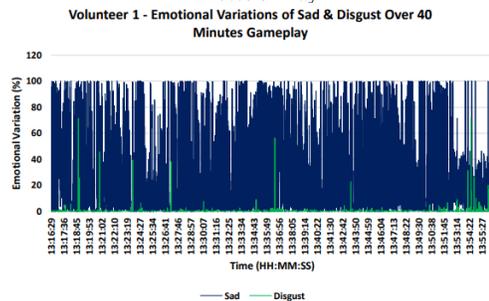

Fig. 7. Volunteer 1: Sad and Disgust Emotional Variations During 40 Minutes of Play

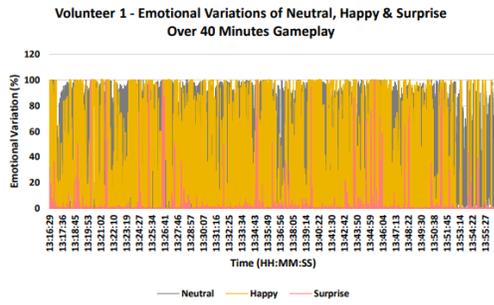

Fig. 8. Volunteer 1: Neutral, Happy, and Surprise Emotional Variations During 40 Minutes of Play

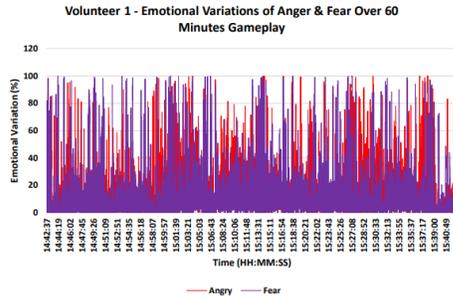

Fig. 9. Volunteer 1: Anger and Fear's Emotional Variations During 60 Minutes of Play

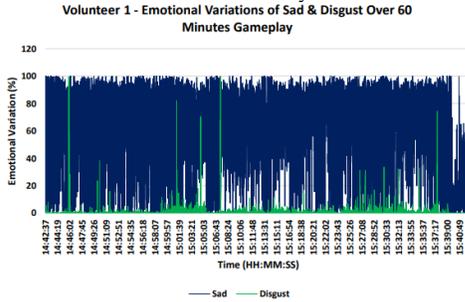

Fig. 10. Volunteer 1: Sad and Disgust Emotional Variations During 60 Minutes of Play

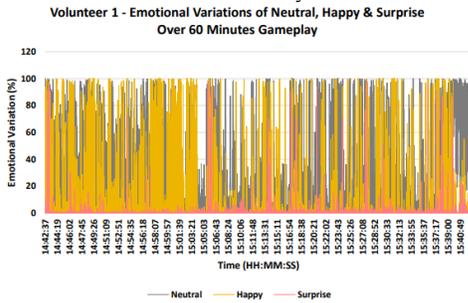

Fig. 11. Volunteer 1: Neutral, Happy, and Surprise Emotional Variations During 60 Minutes of Play

gaming, implying that personal distinctions are crucial in shaping the emotional experiences of gamers.

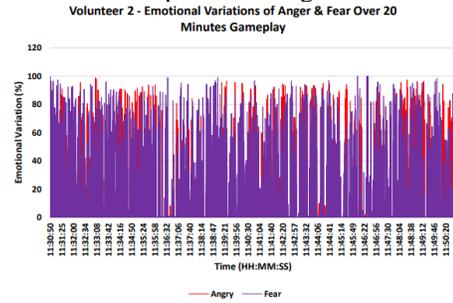

Fig. 12. Volunteer 2: Anger and Fear's Emotional Variations During 20 Minutes of Play

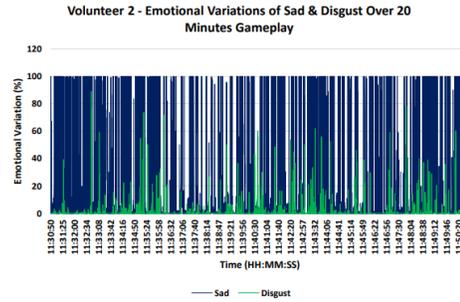

Fig. 13. Volunteer 2: Sad and Disgust Emotional Variations During 20 Minutes of Play

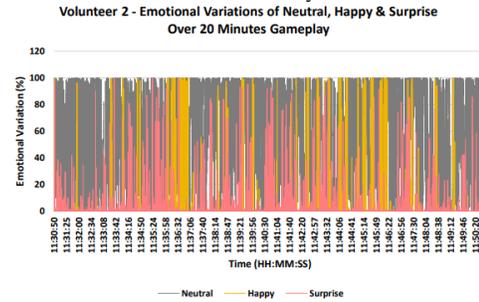

Fig. 14. Volunteer 2: Neutral, Happy, and Surprise Emotional Variations During 20 Minutes of Play

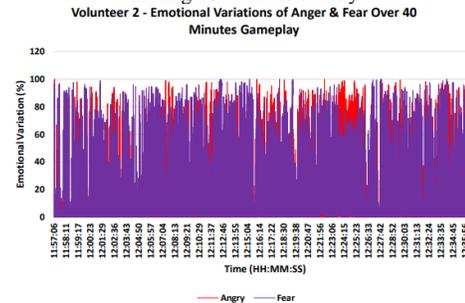

Fig. 15. Volunteer 2: Anger and Fear's Emotional Variations During 40 Minutes of Play

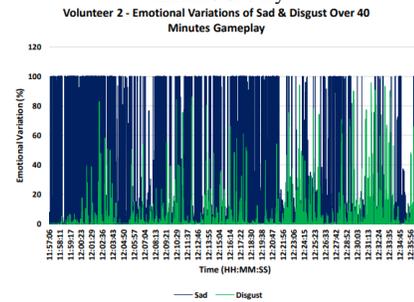

The study also brought attention to the variations in emotional reactions among individuals. During the 20-minute session (refer to Figs. 12, 13, and 14), the second participant, for example, felt a more balanced distribution of anger and fear along with notable amounts of grief and disgust. In contrast to the first volunteer's experience, neutrality was strongly experienced. As the sessions extended to 40 (refer to Figs. 15, 16, and 17) and 60 minutes (refer to Figs. 18, 19, and 20), there was a shift in the emotional patterns, with grief and contempt displaying varying degrees of intensity and fear and wrath becoming more prominent. The aforementioned diversity highlights the personalised aspect of emotional reactions to

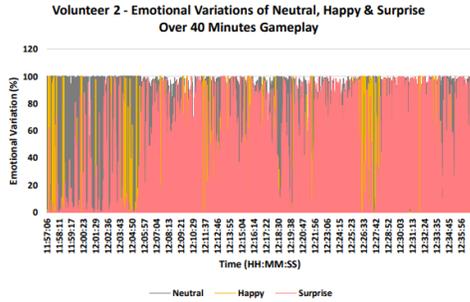

Fig. 16. Volunteer 2: Sad and Disgust Emotional Variations During 40 Minutes of Play

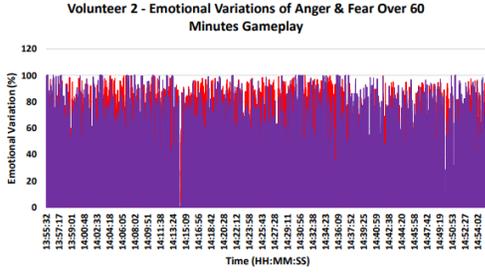

Fig. 17. Volunteer 2: Neutral, Happy, and Surprise Emotional Variations During 40 Minutes of Play

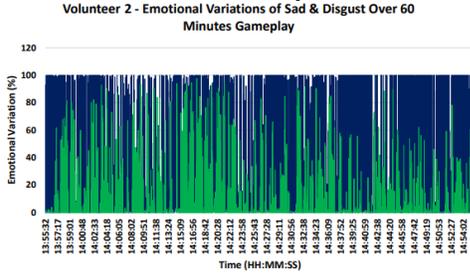

Fig. 18. Volunteer 2: Anger and Fear's Emotional Variations During 60 Minutes of Play

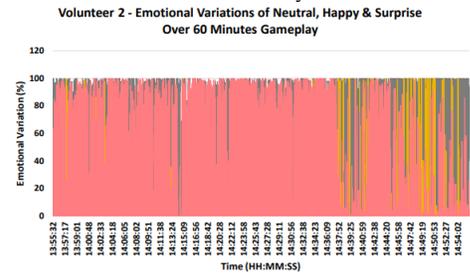

Fig. 19. Volunteer 2: Sad and Disgust Emotional Variations During 60 Minutes of Play

Fig. 20. Volunteer 2: Neutral, Happy, and Surprise Emotional Variations During 60 Minutes of Play

With the help of quantitative data and subjective evaluations, the study looked at players' feelings during various durations of gaming sessions to provide a complete picture of the emotional terrain of digital gaming. At first, especially in the shorter sessions, players experienced enthusiasm and attention. But after an hour-long session, participants reported feeling tired and losing interest, which coincided with a rise in negative emotions detected by emotion recognition software. The psychological effects of extended gaming on players' emotional states are highlighted in this thorough investigation, which provides insightful information for game designers and developers. By examining the intricate emotional dynamics players encounter, it strengthens our understanding of human-computer interaction and emphasizes the importance of taking mental health into account when designing captivating gaming experiences.

## V. RESULTS ANALYSIS

Peoples choose the game they want to play in real life, but they aren't able to fully see how the gameplay affects their feelings. As a result, these players don't think to express their feelings in their reviews to the corporations that make and design the games. Therefore, it becomes harder for the companies who make and design these games to come up with new, improved, and thrilling games quickly enough to meet the demand. This led to the decision that the goal should be to determine how the player's emotions are impacted by the duration of the activity. Based on the conducted experiment, the findings show that the participants' emotions tended to become less joyful and more neutral as the gameplay duration extended from 20 to 60 minutes. During the 20-minute experiment, both individuals consistently displayed higher degrees of dread. This implied that these volunteers might have been more afraid due to the intensely competitive atmosphere the game created and their concern that opponents would shoot them from any angle. This would have enhanced their senses because they would have been concentrating on keeping an eye on their surroundings while playing. Similar findings were made by [17] in her study, which showed that playing action video games increased stress levels compared to playing non-action games. The second participant felt the strongest emotions—anger, contempt, neutrality, and surprise—of all the others. This is similar to a study by [14], which discovered that those who played the violent game online were more irrational and violent than people who played it offline. But compared to the second volunteer, this one felt far less rage and disdain. This could imply that, in comparison to the second volunteer, the first volunteer wasn't as concerned about dying constantly during playtime. This study is similar to [18] in that it discovered that, despite the game's content, participants who had been initially irritated later released their negative feelings and reported that the gameplay was more enjoyable. One possible explanation for the low degrees of anger and contempt could be because the second volunteer is already accustomed to the intense emotional content of first-person shooter games. The results of the study showed that during gaming sessions, volunteers felt less angry and more satisfied, with the lowest levels of despair and significantly higher levels of happiness. This is consistent with research [14], where participants relished a violent game in spite of its difficulties. More powerful emotions were evoked during the 40-minute sessions, probably as a result of the greater focus and effort. In line with [15]'s findings that violent content and gameplay errors can intensify feelings of rage and frustration, one participant displayed similar emotions throughout sessions, with increased irritation following gaming errors. This suggests that getting too comfortable with a game's difficulty may lessen excitement. Extended gaming sessions were associated with moderate increases in disdain and sadness, but volunteers reported feeling content despite distractions. According to research, playing in multiplayer modes has been demonstrated to lessen rage [19]. The violent aspect of the game was linked to the rising wrath over a 60-minute period, which was consistent with findings by [14]. Anomalies in emotion recognition highlighted possible data lag concerns.

## VI. Conclusion and Future Works

The association between the emotions felt during gameplay and the gameplay itself is confirmed by experiments conducted on earlier work. The purpose of this study was to investigate how a player's emotions are influenced by the duration of games. The results showed that from the 20-minute to the 60-minute session, there was a discernible rise in the magnitude of the emotions felt. These results, along with the subsequent research, lead to the conclusion that, in comparison to shorter gameplay sessions, the longer gameplay sessions did have a substantial effect on the player's emotions. Longer gameplay had an impact on all the emotions, but the negative emotions—such as anger and sadness—showed a notable rise. In light of these results, to lessen the potential emotional harm that playing computer and video games may cause in the future, game creators should use these findings to design and develop more thrilling games that can be finished faster. One approach that offers a promising avenue for improving player emotional engagement is the integration of dialogue systems [20] and audio fingerprinting [21]. By adapting in real time to the player's emotional state, these technologies can improve the quality of gameplay by adjusting in-game audio and NPC interactions. Future works will focus on utilizing these instruments to alleviate adverse affective states during prolonged gameplay, thereby potentially converting gaming into a more emotionally suitable and immersive encounter. In addition, further investigation should be conducted to determine the ideal amount of time for players to interact with the game, allowing them to enjoy it with the least amount of negative emotional impact.

## VII. Declarations

*A. Funding:* No funds, grants, or other support was received.

*B. Conflict of Interest:* The authors declare that they have no known competing financial interests.

*C. Data Availability:* Data will be made on reasonable request.

*D. Code Availability: Code will be made on reasonable request.*